\documentclass[twocolumn]{aastex63}

\usepackage{etoolbox}
\makeatletter 
 \patchcmd{\NAT@citex}
   {\@citea\NAT@hyper@{%
     \NAT@nmfmt{\NAT@nm}%
     \hyper@natlinkbreak{\NAT@aysep\NAT@spacechar}{\@citeb\@extra@b@citeb}%
     \NAT@date}}
   {\@citea\NAT@nmfmt{\NAT@nm}%
   \NAT@aysep\NAT@spacechar\NAT@hyper@{\NAT@date}}{}{}

 \patchcmd{\NAT@citex}
   {\@citea\NAT@hyper@{%
     \NAT@nmfmt{\NAT@nm}%
     \hyper@natlinkbreak{\NAT@spacechar\NAT@@open\if*#1*\else#1\NAT@spacechar\fi}%
       {\@citeb\@extra@b@citeb}%
     \NAT@date}}
   {\@citea\NAT@nmfmt{\NAT@nm}%
   \NAT@spacechar\NAT@@open\if*#1*\else#1\NAT@spacechar\fi\NAT@hyper@{\NAT@date}}
   {}{}
\makeatother

\newcommand{\pmn}{\mathcal{M}_{\Psi}}

\newtoggle{referee}            %
\togglefalse{referee}          %

\iftoggle{referee}{
  \usepackage{color,soul}    %
  \soulregister\citet7
  \soulregister\citep7
  \soulregister\ref7
  \newcommand{\hlm}[1]{%
      \hl{#1}
  }                          %
}{
  \newcommand{\hlm}[1]{{#1}} %
}                              %

\usepackage{xcolor}
\usepackage{amsmath}
\usepackage{CJK}

\received{July 10, 2020}
\revised{August 12, 2020}
\accepted{August 30, 2020}

\submitjournal{ApJ}

\shorttitle{Diffusive Radiation Transport in Turbulent Plasma}
\shortauthors{Schultz, Bildsten, and Jiang}

\graphicspath{{./}{figures/}}

\begin{document}
\begin{CJK*}{UTF8}{gbsn}

\title{Convectively Driven Three Dimensional Turbulence in Massive Star Envelopes: \\ I. A 1D Implementation of Diffusive Radiative Transport \footnote{Released on ??, ??, 2020}}

\correspondingauthor{W. C. Schultz}
\email{wcschultz@physics.ucsb.edu}

\author[0000-0002-0786-7307]{William C. Schultz}
\affiliation{Department of Physics, University of California, Santa Barbara, CA 93106, USA}

\author{Lars Bildsten}
\affiliation{Department of Physics, University of California, Santa Barbara, CA 93106, USA}
\affiliation{Kavli Institute for Theoretical Physics, University of California, Santa Barbara, CA 93106, USA}

\author[0000-0002-2624-3399]{Yan-Fei Jiang(姜燕飞)}
\affiliation{Center for Computational Astrophysics, Flatiron Institute, New York, NY 10010, USA}



\begin{abstract}
Massive ($M>30\,$M$_{\odot}$) stars exhibit luminosities that are near the Eddington-limit for electron scattering causing the increase in opacity associated with iron at T$\,\approx 180,000\,$K to trigger supersonic convection in their outer envelopes. 
Three dimensional radiative hydrodynamics simulations by Jiang and collaborators with the Athena++ computational tool have found order of magnitude density and radiative flux fluctuations in these convective regions, even at optical depths $\gg100$. 
We show here that radiation can diffuse out of a parcel during the timescale of convection in these optically thick parts of the star, motivating our use of a ``pseudo" Mach number to characterize both the fluctuation amplitudes and their correlations.  
In this first paper, we derive the impact of these fluctuations on the radiative pressure gradient needed to carry a given radiative luminosity.  
This implementation leads to a remarkable improvement between 1D and 3D radiative pressure gradients, and builds confidence in our path to an eventual 1D implementation of these intrinsically 3D envelopes. However, simply reducing the radiation pressure gradient is not enough to implement a new 1D model.  
Rather, we must also account for the impact of two other aspects of turbulent convection: the substantial pressure, and the ability to transport an appreciable fraction of the luminosity, which will be addressed in upcoming works. 
This turbulent convection also arises in other instances where the stellar luminosity approaches the Eddington luminosity. 
Hence, our effort should apply to other astrophysical situations where an opacity peak arises in a near Eddington limited, radiation pressure dominated plasma.
\end{abstract}

\keywords{Stellar convective shells (300), Stellar processes (1623), Stellar photospheres (1237), Stellar surfaces(1632), Stellar structures(1631), Stellar physics(1621), Stellar convection envelopes(299)}

\section{Introduction} \label{sec:intro}
Massive stars play a fundamental role in many astrophysical environments.
The ionizing radiation from massive stars was important in the reionization of the early universe \citep{Bromm2004}.
This radiation, along with strong winds produced by massive stars, are important feedback mechanisms that regulate star formation and the structure of the interstellar medium in galaxies \citep{Kennicutt2005, Smith2014}.
The explosions of massive stars produce various types of supernovae and high energy transients and leave behind black holes and neutron stars.
The properties of both the explosions as well as the remnants depend strongly on the evolution of the massive star progenitor \citep{Heger2003, Farmer2016}. 

However, modeling massive stars is notoriously challenging.
The fundamentals of hydrostatics and radiative heat transport in massive stars ($M>50-100\,$M$_{\odot}$) cause a luminosity that approaches the Eddington-limit for electron scattering as well as an increasing dominance of radiation pressure over that of the gas \citep{Crowther2007, Maeder2012, Sanyal2015}. 
For stars with near solar metallicity, the opacity increase associated with iron at $T\,\approx 1.8\times 10^5\,$K implies a locally super-Eddington luminosity  \citep[e.g.][and references therein]{Paxton2013}.
Recent 3D calculations \citep{Jiang2015, Jiang2018} found surprising properties of these massive star envelopes.
For example, while a deep 1D convective zone has no direct impact on the surface properties, the 3D calculations revealed that velocity and density fluctuations propagate well into the stellar photosphere \citep{Jiang2015, Jiang2018}.
These simulations also found a complex interplay of convective and radiative transport whose behavior depends on the ratio of the photon diffusion time to the local dynamical time.
Even further out, \citet{Jiang2018} showed that helium recombination causes an even larger increase in opacity that can lead to a continuum driven wind.
Though these 3D models provide important information regarding the dynamics in massive star envelopes, they take over ten million CPU hours to run for a month of model time.
Hence in order to see how massive stars evolve throughout their lives, accurate 1D modeling is required.

In these radiation pressure dominated regions with super-Eddingtom luminosities, one dimensional stellar models yield density and gas pressure inversions \citep{Joss1973, Grafener2012, Paxton2013, Owocki2015} that trigger new convective instabilities. 
These instabilities stymie 1D models when the iron opacity peak is close to the surface where mixing length theory (MLT; \citealt{Cox1968}) convection is inefficient and cannot carry the necessary luminosity.
In the absence of an improved 1D modeling approach, stellar modelers simply bypass this obstacle with explicit ``patches'' that enable continued evolution \citep[e.g.][]{Stothers1979,Maeder1987,Paxton2013}.
However, these ``patches'' have not yet been tested for physical accuracy and a physically motivated solution is highly desired \citep[e.g. Figure 20 of ][]{Kohler2015}. 

In this paper, we present a new prescription to calculate radiation pressure gradients in the outer layers of 1D models of massive stars.
Our prescription incorporates the correlations between fluctuations of density and radiative flux observed in the 3D models of \citet{Jiang2018}.
As these stars are radiation pressure dominated, accurately determining the radiation pressure gradient is fundamental to the structure of their stellar envelopes throughout the star's evolution.
In developing the prescription, we also identified a useful independent parameter to identify regions of inefficient convection across different 3D models with different masses and luminosities.

Strong effects caused by the correlation between density and radiative flux in optically thin, $\tau \lesssim 1$, have been detailed in previous literature.
Many investigations that considered the interaction between density fluctuations and radiation in the context of radiation-driven turbulent winds in star forming regions, finding significant effects from their correlation \citep[e.g.][]{Krumholz2012, Davis2014, Rosdahl2015, Tsang2015}.
\citet{Krumholz2013} noted the pronounced effect this correlation has on estimations of the asymptotic momentum of dusty, radiative driven winds.
\citet{Tsang2018} observe a sizable effect of the correlation between radiative flux and density in the super star cluster formation and evolution.
Recently, many advances have been made towards characterizing the effect of density fluctuations on the radiation transport through clumpy stellar winds \citep[e.g.][and references therein]{Owocki2018}.
Despite these correlations being observed, no work has been able to reproduce the effect of these correlations in either 1D evolution or subgrid recipe contexts.
Though these correlations have been identified in other astrophysical environments, proper analysis has not been applied to the optically thick, $\tau \gg 1$, regions of massive star envelopes.

The remainder of the paper is organized as follows.
In Section~\ref{sec:3D_models} we detail the 3D models used in our analysis, characterizing fluctuations in density and radiative flux, and quantifying their variance and covariance.
We also introduce a new independent parameter that characterizes the turbulent convection.
Section~\ref{sec:psi} describes our proposed prescription as well as providing evidence for its efficacy.
We discuss the implications of our prescription in future 1D stellar evolution models in Section~\ref{sec:1D_check}, and summarize our key results in Section~\ref{sec:conclusion}.

\begin{deluxetable*}{lccccccccccc}
\tablenum{1}
\tablecaption{Properties of the 3D Stellar Models \label{tab:3D_models}}
\tablewidth{0pt}
\tablehead{
\colhead{Model Name} & \multicolumn{2}{c}{Masses} &\colhead{Temperature} & \multicolumn{2}{c}{Luminosities} & \multicolumn{4}{c}{Radii} & \colhead{Optical Depth} & \colhead{Metallicity} \\
\nocolhead{Name} & \colhead{$M_{\rm core}$} & \colhead{$M_{\rm env}$} & \colhead{$T_{\rm eff}$} & \colhead{$L$} & \colhead{$L_{\rm Edd}$ $^{\rm a}$} & \colhead{$r_{\rm base}$} & \colhead{$r_{\rm max}$} & \colhead{$r_{\rm ph}$ $^{\rm b}$} & \colhead{$r_{\rm Fe }$} & \colhead{$\tau_{\rm Fe}$} & \colhead{$Z$} \\
\nocolhead{Name} & \colhead{(M$_{\odot}$)} & \colhead{(M$_{\odot}$)} & \colhead{($10^3\,$K)} & \colhead{(log($L$/L$_{\odot}$))} & \colhead{(log($L$/L$_{\odot}$))} & \colhead{(R$_{\odot}$)} & \colhead{(R$_{\odot}$)} & \colhead{(R$_{\odot}$)} & \colhead{(R$_{\odot}$)} & \nocolhead{tau} & \colhead{(Z$_{\odot}$)}
}
\startdata
T9L6.2 & 56 & 0.13 & 9 & 6.20 & 6.26 & 35.0 & 809.8 & 353.3  & 80.3 & 28,000 & 1 \\
T19L6.4 & 80 & 0.011 & 19 & 6.40 & 6.42 & 16.3 & 335.5 & 99.0 & 44.0 & 5,400 & 1 \\
T19L6.0 & 35 & 0.0002 & 19 & 6.00 & 6.06 & 18.2 & 387.1 & 71.1 & 25.2 & 4,100 & 1 \\
T9L6.2Z0.1 & 56 & 0.13 & 9 & 6.20 & 6.26 & 35.0 & 809.8 & 322.5 & 87.5 & 16,000 & 0.1 \\
T19L6.4Z0.1 & 80 & 0.011 & 19 & 6.40 & 6.42 & 16.3 & 335.5 & 102.2 & 45.7 & 3,100 & 0.1 \\
T19L6.4Z2 & 80 & 0.011 & 19 & 6.40 & 6.42 & 16.3 & 335.5 & 104.7 & 45.7 & 8,300 & 2 \\
\enddata
\tablenotetext{\rm a}{ For an assumed electron scattering opacity.}
\tablenotetext{\rm b}{ The photosphere radii specified are where $\langle \tau(r_{\rm ph}) \rangle = 1$. (see equation~(\ref{eq:ang_avg}))}
\end{deluxetable*}

\section{3D Simulations of Massive Star Envelopes} \label{sec:3D_models}
The 3D simulations used in this work modeled the outer $<1\%$ of massive star envelopes with the radiation MHD code {\sf Athena++} \citep{Stone2020}.
The code solves the ideal hydrodynamic equations coupled with the time-dependent, frequency-integrated radiation transport equation for specific intensities over discrete angles based on the numerical algorithm as described by \cite{Jiang2014}. 
All the simulations are done in the spherical polar coordinate with effective resolution $512\times 512\times 256$ covering the radial, longitudinal $(\phi\in[0,\pi])$ and latitudinal ($\theta\in [\pi/4,3\pi/4]$) directions. 
The angular grid for the radiation field are constructed in the same way as specified in \cite{Davis2012}. 
Results for three simulations were briefly described by \cite{Jiang2018} and are being additionally studied in a paper in preparation. 
Here we include a few additional models with different metallicities. 
The physical characteristics of the models are listed in Table~\ref{tab:3D_models}. These include the core mass beneath the model, $M_{\rm core}$, the envelope mass being simulated, $M_{\rm env}$, the effective temperature, $T_{\rm eff}$, the luminosity, $L$, the Eddington luminosity for electron scattering, $L_{\rm Edd} = 4\pi G M_{\rm core} c / \kappa_{\rm es}$, and the radial location of the bottom, $r_{\rm base}$, and the top, $r_{\rm max}$, of the model grids. 
We also show the radial location where the expected value of optical depth to infinity is unity, $r_{\rm ph}$, the radial location and optical depth at the iron opacity peak, $r_{\rm Fe}$ and $\tau_{\rm Fe}$, and the metallicity, $Z$.
We focus here on the 3D model properties when they have reached a steady-state equilibrium, allowing us to assume that optically thick regions have reached local thermal equilibrium.
Despite the models reaching steady-state they still exhibit luminosity variations on the order of $50\%$ on time scales of days \citep{Jiang2018}.
These envelope models only account for the gravitational accelerations caused by the cores, which account $>99\%$ of the stellar mass.
The models do not include any initial rotation.
In this paper, we use model T9L6.2 to investigate the properties of the turbulent envelopes (Section~\ref{sec:rad_var}), models T9L6.2 and T19L6.4 to quantify the variations in density and radiative flux (Section~\ref{sec:sig_cov}), and the remaining four models (T19L6.0, T9L6.2Z0.1, T19L6.4Z0.1, T19L6.4Z2) to test the prescription we developed (Section~\ref{sec:1D_check}).

Throughout these envelope models, the radiation pressure is substantially larger than the gas pressure by nearly a factor of 10. 
However, as the optical depth decreases at larger radii, the photons can no longer support the gas against local compression and the gas pressure becomes dominant post-shock support. 
This occurs at a critical optical depth defined as $\tau_{\rm c} \approx c/c_{\rm g,0}$ where $c_{\rm g,0}$ is the isothermal gas sound speed at the iron opacity peak \citep{Jiang2015}.
The iron opacity peak is a convenient choice as it typically instigates the majority of the convection in massive star envelopes and occurs at the same temperature so $\tau_{\rm c} \approx 6,300$ for all the models.
In our models, the iron opacity peak causes the luminosity to become super-Eddington, causing turbulence and subsequent density fluctuations as we shall discuss.
\citet{Jiang2015} highlighted the contrast in the outcome of convective properties as a function of $\tau$.
If convection occurs at $\tau \ll \tau_{\rm c}$, the plasma is optically thick enough to be locally supported by the radiation pressure allowing the convection to be efficient and reasonably well described by classical MLT.
If convection occurs at $\tau \gg \tau_{\rm c}$, the plasma is optically leaky, letting the photons escape and leaving only the minimal gas pressure to support against local perturbations.
This convection will be inefficient and may behave very differently when compared to classical MLT. 
The models in Table~\ref{tab:3D_models} span this boundary at the iron opacity peak and, as we show, exhibit large density fluctuations as $\tau \ll \tau_{\rm c}$.

\subsection{3D Radial Variations}\label{sec:rad_var}
One distinctive characteristic of these 3D massive star envelope simulations is the extraordinary variations in density, opacity, optical depth, and radiation flux deep in the models, at $\tau \gg 1$. 
At the base, where the models are entirely radiative, the opacity is dominated by electron scattering, and the luminosity is very close to the local Eddington luminosity.
As the temperature decreases outwards, the plasma cools enough for iron to cause an increase in opacity once the temperature reaches $T\approx 1.8\times 10^5\,$K, known as the iron opacity peak. 
This increase in opacity decreases the local Eddington luminosity causing the stellar luminosity to surpass it, which results in vigorous convection \citep{Joss1973, Paxton2013} that causes large density fluctuations in this region. 
Even further out, helium recombination causes an even larger increase in opacity; the helium opacity peak.
This opacity peak is sensitive to density which, combined with the large density fluctuations of the turbulent convection, leads to radiation accelerating dense clumps.
Some clumps reach the escape velocity, while most others fall back onto the star \citep{Jiang2018}.
The cyclic motion of this process creates large inhomogeneities above the surface of the star and large optical depth fluctuations.

\begin{figure}
\centering
\includegraphics[width=\linewidth]{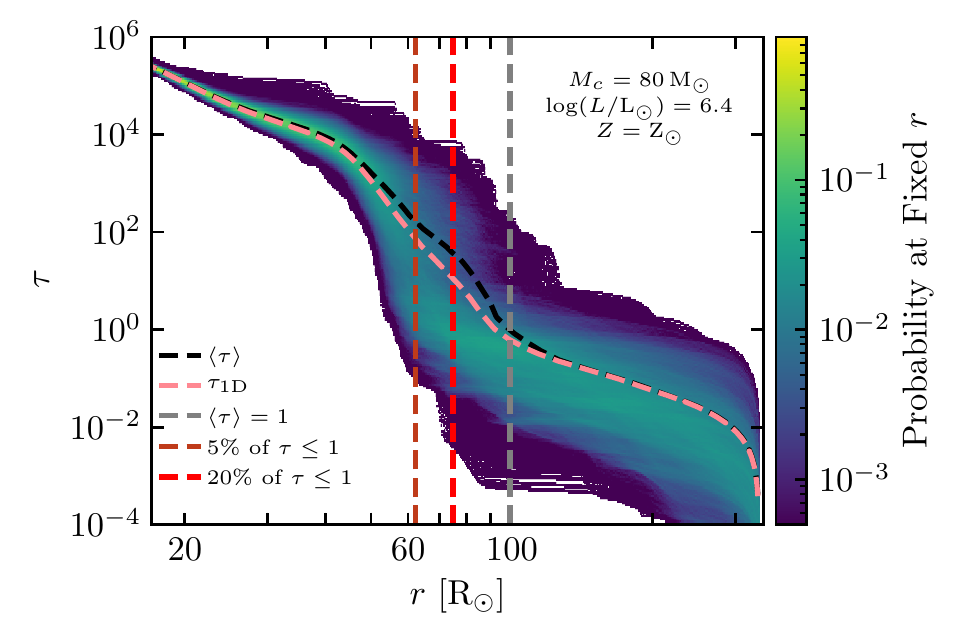}
\caption{Probability mass distribution of radial optical depth, $\tau$, to infinity as a function of radius for one snapshot of model T19L6.4. 
The optical depth is calculated by integrating radially along lines of constant angle. 
The black-dashed line shows the angular average of $\tau$, weighted by the cell's area. 
The grey vertical dashed line shows where $\langle \tau \rangle = 1$, a typical definition of the photosphere. 
The pink-dashed line shows the radial profile of a 1D approximation of $\tau$, $\tau_{1D}$ (defined in equation~(\ref{eq:tau_1D})).
We denote the radii where $5\%$ (dark red-dashed) and $20\%$ (bright red-dashed) of the shell's area has an optical depth to infinity of 1 or less. 
Each distribution is generated from optical depths calculated in the 131,072 cells at each radius. 
\label{fig:tau_v_r}}
\end{figure}

To quantify these variations, we first calculate the optical depth of a given location in the models by integrating from the location's radius to the edge of the simulation along the radial direction. 
The result is one $\tau$ measurement per cell or 131,072 measurements per radial shell, as set by the resolution of these simulations.
For diagnostic purposes, the radial integral is adequate and so we do not calculate $\tau$ for other directions.
Figure~\ref{fig:tau_v_r} shows the probability distributions of $\tau$ as a function of radius for a single snapshot of the T19L6.4 model. 
The color represents the probability of finding a specific value of $\tau$ at a given radius. 
The vertical axis is logarithmic in scale and in certain regions, the optical depth spans six orders of magnitude!
The black-dashed line is the radial average of $\tau$ weighted by each cell's radial area , $\langle \tau(r) \rangle$, defined by,
\begin{equation} \label{eq:ang_avg}
\langle \tau(r) \rangle \equiv \frac{\int^{\phi_{\rm max}}_{\phi_{\rm min }}\int^{\theta_{\rm max}}_{\theta_{\rm min}} \tau(r, \theta, \phi) d(\cos \theta) d\phi}{\int^{\phi_{\rm max}}_{\phi_{\rm min}}\int^{\theta_{\rm max}}_{\theta_{\rm min}} d(\cos \theta) d\phi},
\end{equation}
where the angle integrals are evaluated over the entire solid angle of the simulation.
Angle brackets will represent angular averages at fixed radius for the remainder of the work.
The pink line is the optical depth calculated from the average opacity and average density, or 
\begin{equation} \label{eq:tau_1D}
    \tau_{\rm 1D}(r) \equiv \int^{r_{\rm max}}_{r} \langle \kappa(r') \rangle \langle \rho(r') \rangle dr' ,
\end{equation}
where $r_{\rm max}$ is the outer radius of the simulation grid, listed in Table~\ref{tab:3D_models}.
The two red, vertical dashed lines represent the radius where $5\%$ (dark red) and $20\%$ (bright red) of the shell's area have an optical depth of less than 1. 
The percentages were chosen to approximate, to varying degrees, where the assumption of optically thick radiative transport would begin to fail.
These locations are significantly different from the location of the photosphere as defined by either $\langle \tau \rangle = 1$ or $\tau_{\rm 1D} = 1$. 

\begin{figure*}[ht!]
\centering
\gridline{\fig{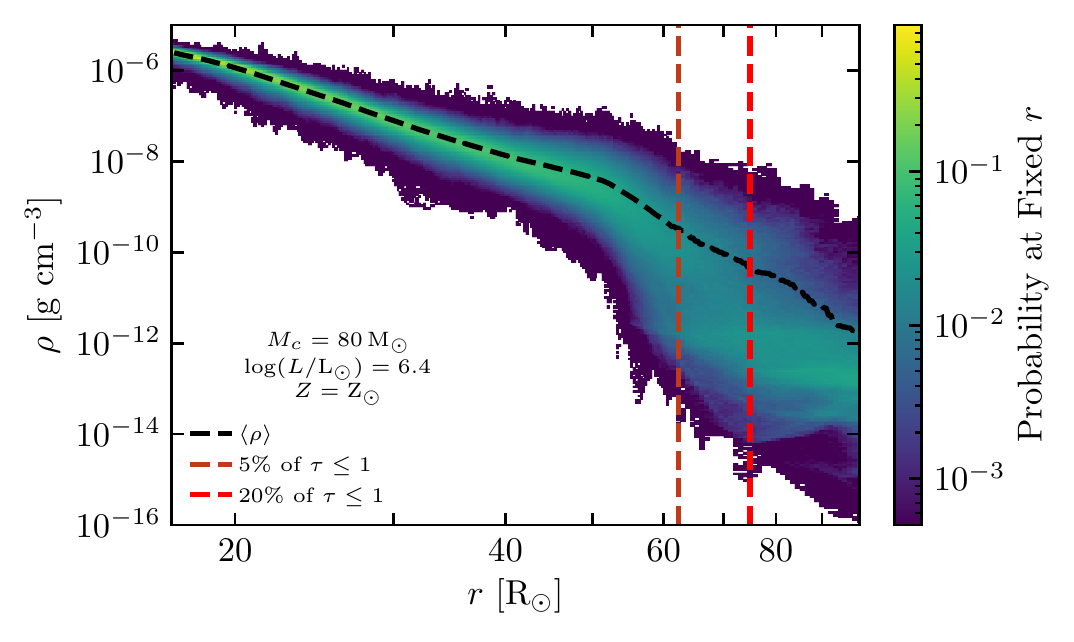}{0.45\linewidth}{(a)}
 \fig{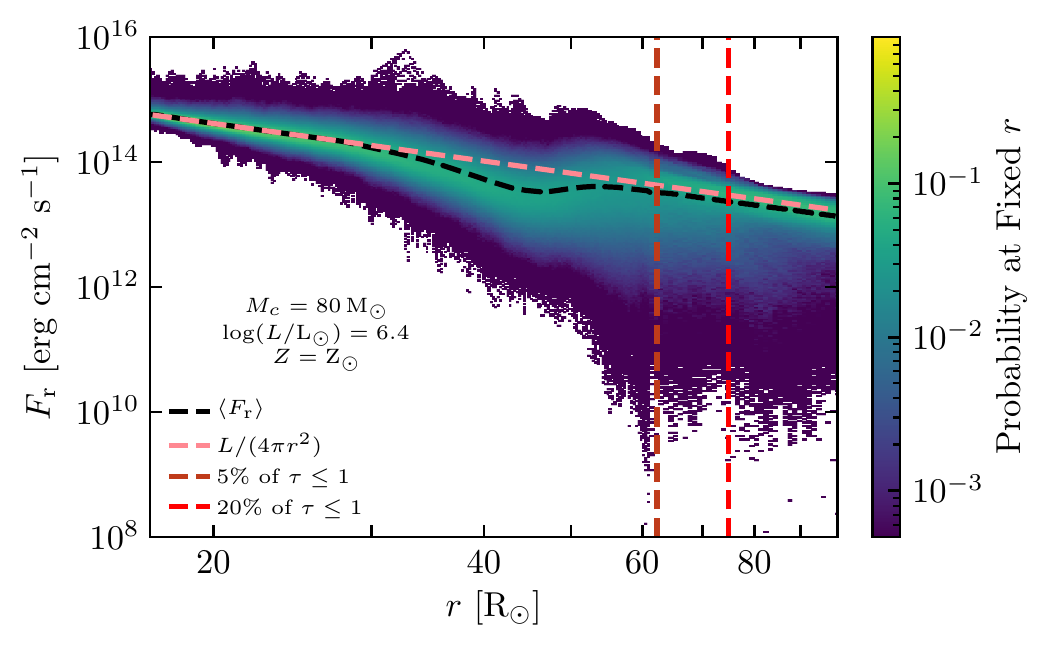}{0.45\linewidth}{(b)}
}
\gridline{\fig{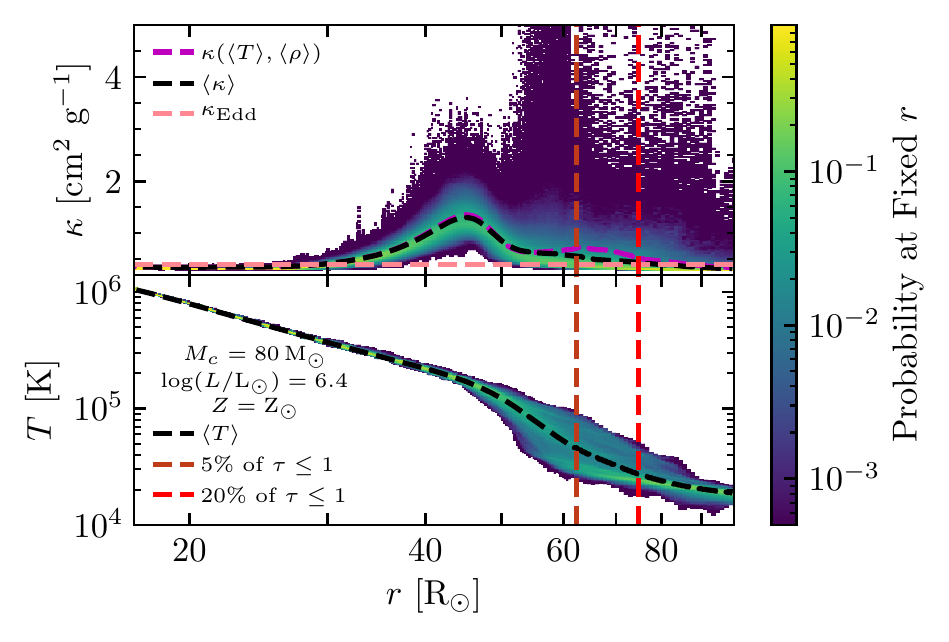}{0.45\linewidth}{(c)}
\begin{minipage}[b]{0.45\textwidth}
\caption{Probability mass distributions of (a) density, (b) radiative flux, (c) opacity and temperature as a function of radius for model T19L6.4.
The black-dashed line represents the average value. 
The vertical red lines are the same as in Figure~\ref{fig:tau_v_r}, showing the radial location where an optically thick assumption would breakdown.
The opacity calculated using the average temperature, $\langle T \rangle$, and average density, $\langle \rho \rangle$, (magenta-dashed line) as well as the opacity needed for a local Eddington ratio of 1 (pink-dashed line) are shown in the upper panel of (c).
The pink-dashed line in (b) represents the total average flux, $L/(4\pi r^2)$, for this snapshot.
The distributions are generated from 131,072 cells at each radius. \label{fig:other_v_r}}
\end{minipage}
}
\end{figure*}

The variations in density, $\rho$, radiative flux, $F_r$, opacity, $\kappa$, and temperature, $T$, are shown in Figure~\ref{fig:other_v_r} for a single snapshot of model T19L6.4. 
Similarly to Figure~\ref{fig:tau_v_r}, the color represents the probability that the quantity has the given value at that particular radius.
The angle average of the quantities, shown by the black-dashed lines, mostly agree with the center of the probability distributions at all optically thick radii.
The temperature has small fluctuations below the iron opacity peak and only varies by a factor of a few above it.
\hlm{Radiation pressure is the main source of hydrostatic support of these outer envelopes, constraining the temperature fluctuations to be relatively low at any fixed radius.} 
The density variations at a fixed radius grow quickly above the iron opacity peak, where the strong turbulent motion initiates, and $\rho$ can vary by as much as six orders of magnitude at a fixed radius. 
The iron opacity peak is also where the convective flux becomes comparable to the radiative flux.
Comparing the average total flux, $L/(4\pi r^2$) (pink line in Figure~\ref{fig:other_v_r}b), to the average radiative flux, $\langle F_r \rangle$ (black line in Figure~\ref{fig:other_v_r}b), shows that the convective flux is the dominant form of energy transport throughout the iron opacity peak.
The turbulent convection continues to carry a significant fraction of the flux throughout the remainder of the optically thick region.
\hlm{The small opacity variations in the optically thick regions are due to the low levels of temperature fluctuations. Though the density variations can be large, the opacity is not very sensitive to density in these locations}.
However in the outer layers, helium recombination increases the opacity (up to 10$\,$cm$^2$/g), giving rise to optically thick winds in these models \citep{Jiang2018}.
The model is also substantially Eddington-limited for much of its radial extent as shown by the pink-dashed line in Figure~\ref{fig:other_v_r}c, which represents the opacity, $\kappa_{\rm Edd}$, required for the local Eddington ratio to be unity.
The mean opacity quickly surpasses it at the iron opacity peak.

\subsection{Correlation of Flux, Opacity, and Density} \label{subsec:correlations}

\begin{figure*}[ht!]
\centering
\includegraphics[width=\linewidth]{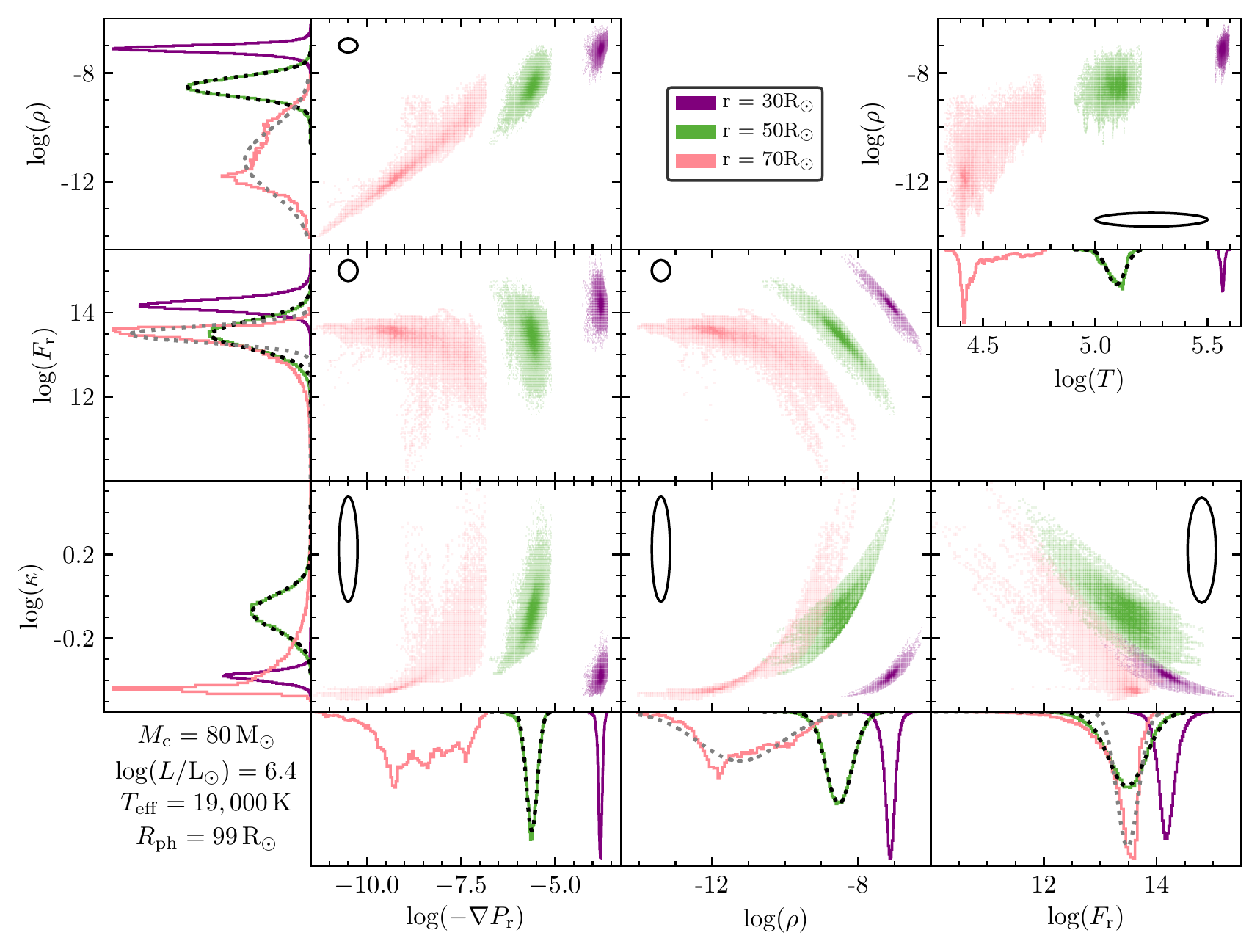}
\caption{Distributions of and correlations between density, radiative flux, opacity, the radiation pressure gradient, and temperature at three radii for model T19L6.4.
The three radii choices are just below the iron opacity peak (purple), just above the iron opacity peak (green), and where the optically thick radiative transfer  assumption begins to break down (pink). 
The black lines are circles with the radii of $\log(x)=0.25$ and highlight the difference in axis scaling.
The black-dotted line in the distributions of density, radiative flux, opacity, and radiation pressure gradient are log-normal distribution fits to the intermediate radius choice (green).
The grey-dotted line in the distributions of density and radiative flux are log-normal distribution fits to the outer radius choice (pink).
Each distribution is generated by sorting the quantities in the 131,072 cells at each radius into 100 bins.
\label{fig:corr}}
\end{figure*}

Figure~\ref{fig:corr} shows the distributions of density, radiative flux, opacity, temperature, and the radiation pressure gradient as well as their correlations at three radii of model T19L6.4. 
Here, and for the rest of the work, the symbol $\nabla P_{\rm r}$ is used for  the radiation pressure gradient as calculated by the radial area weighted average of the diffusion equation,
\begin{equation} \label{eq:Pr_OG}
    \nabla P_{\rm r} \equiv -\frac{1}{c} \langle F_{\rm r} \, \kappa \, \rho \rangle.
\end{equation}
The correlations and distributions of quantities at a radius below the iron opacity peak, $r = 30\;$R$_{\odot}$, are shown by the purple distributions. 
Deep in the stellar envelope, fluctuations of all quantities follow log-normal distributions at a fixed radius and are well represented by the probability distribution function,
\begin{equation}
\begin{aligned} \label{eq:log_norm}
    & f(x) = \frac{1}{\sqrt{2 \pi} \sigma_{\ln(x)}} \exp \left(-\frac{(\ln(x) - \langle \ln(x) \rangle)^2}{2\sigma_{\ln(x)}^2}\right), \\
    & \text{and } \sigma_{\ln(x)}^2 \equiv \frac{\sum_{i=1}^N \Big(\ln(x_i) - \langle \ln(x) \rangle \Big)^2 V_i}{\sum_{i=1}^N V_i},
\end{aligned}
\end{equation}
where $\sigma_{\ln(x)}$ is the volume-weighted standard deviation of $\ln(x)$, $N$ is the number of angular cells at a given radius, $V_i$ is the volume of cell $i$, and $\int^{\infty}_{-\infty} f(x) d(\ln(x)) = 1$. 
At $r=30\,$R$_{\odot}$, the temperature and opacity fluctuate by less than $25\%$, and the opacity is very similar to $\kappa_{\rm es}$.
In contrast, the density and radiative flux vary by a factor of 10 due to convective undershooting from the turbulent motion above.
The density and radiative flux are also highly inversely correlated, combining with their large fluctuations to cause variations in the radiation pressure gradient.
The opacity increases with density and decreases with radiative flux.
These correlations match intuition: denser regions are more opaque, making it more difficult for radiation to flow through, and vice versa.
There are no strong correlations between $\nabla P_{\rm r}$ and $\rho$, $\kappa$, or $F_{\rm r}$ deep in the envelope.

Moving outward, the next radius choice is just above the iron opacity peak at $r = 50\,$R$_{\odot}$ and is shown by the green distributions.
The fluctuations in all quantities \hlm{are} larger than at $r=30\,$R$_{\odot}$ as turbulent convection is playing an important role in both pressure support and energy transport at $r=50\,$R$_{\odot}$.
All the plotted distributions remain log-normal as shown by the log-normal fits (black-dotted lines in histograms), though temperature is starting to deviate.
The fluctuations of opacity, temperature, and the radiation pressure gradient are still small compared to \hlm{those} of the density or radiative flux.
The radiative flux and density are still highly inversely correlated and density and opacity are now slightly positively correlated with the radiation pressure gradient.
Radiative flux and the radiation pressure gradient are slightly anti-correlated causing the fluctuations in $\nabla P_{\rm r}$ to be smaller than that of either $\rho$ or $F_{\rm r}$.
\hlm{At $r=50\,$R$_{\odot}$, the convective motions behave similar to classical MLT, with hot plasma mixing upwards through cooler regions.
The main difference being the turbulent velocities are trans-sonic causing the large density and thus radiative flux variations seen in the models.}

The pink distributions show quantities at $r=70\,$R$_{\odot}$ where more than $5\%$ of the area has an optical depth to infinity of less than unity. 
The radiation pressure gradient now spans five orders of magnitude, comparable to the variations in density.
The two are now highly correlated, while the radiative flux has become nearly constant, and is approaching the free-streaming radiative flux, $F_{\rm r} \sim E_{\rm r} c$ due to the low optical depths.
The densest cells are still highly correlated with $F_{\rm r}$, however once the radiative flux approaches the free streaming limit, it is no longer correlated with density.
Though the opacity and temperature are varying substantially compared to interior radii, they are still nearly constant when compared to density, radiative flux, and the radiation pressure gradient.
Fluctuations no longer strongly follow log-normal distributions.
\hlm{This is likely due to the nature of convection changing. 
Fluctuations are no longer nearly isotropic, but rather dominated by large, hot, and dense plumes pushing upwards and cool diffuse plasma moving down.
We believe this region to encompass the variable, dynamic surface of these envelopes. The explicit nature of this regions and the convective transport in both regions will be explored further in future work.}
However, as shown by the grey-dotted line in Figure \ref{fig:corr}, distributions of both density and radiative flux variations are still close to log-normal, so we assume both quantities follow log-normal distributions throughout the optically thick regions of these envelope models in order to quantify their variance.

At all radii, temperature and opacity are relatively constant when compared to radiative flux and density and will be treated as such for the remainder of this work, i.e. $\kappa(r) = \langle \kappa(r) \rangle$ and $T(r) = \langle T(r) \rangle$.
\hlm{Including the variations only complicates the analysis and does not have a substantial effect on the resulting prescription.}
The extreme variations in density and radiative flux throughout the stellar envelope combined with their strong inverse correlation strongly modify the radiation pressure gradient, which will be discussed in detail in Section~\ref{sec:psi}.

\subsection{Pseudo-Mach Number Definition and Value} \label{subsec:pseudo_mach}
\begin{figure}[t!]
\centering
\includegraphics[width=\linewidth]{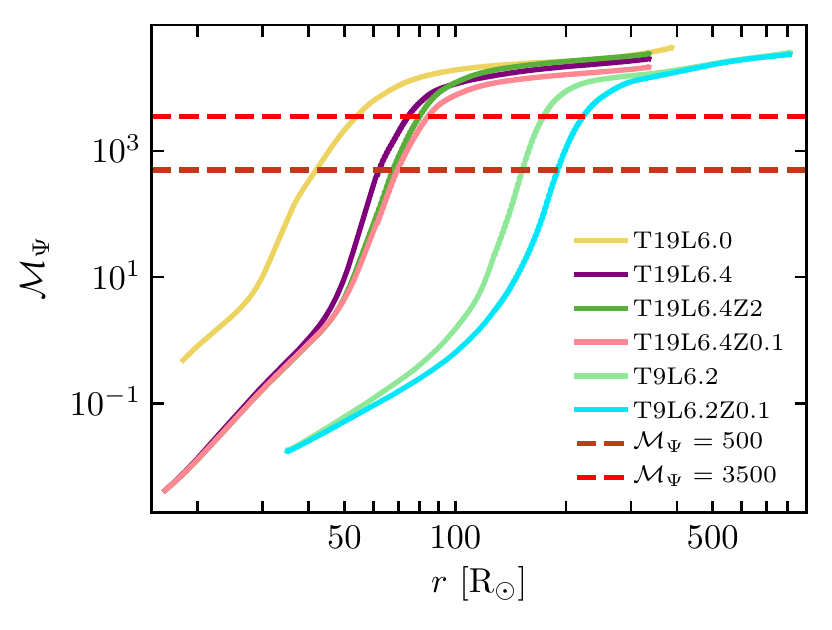}
\caption{Pseudo-Mach number as a function of radius for a single snapshot of all models considered. The horizontal dashed lines correspond to $\pmn= 500$ (dark red) and $\pmn = 3500$ (bright red). \label{fig:pmn_v_r}}
\end{figure}

Ideally, a single, local parameter would quantify the large variations caused by the turbulent motion.
In the classical model of MLT for gas pressure dominated regimes, a turbulent Mach number proves a good choice for an independent variable. 
However, our models are radiation pressure dominated so a new definition is needed. 
Our heuristic choice is motivated by the work of \citet{Jiang2015} and defines a local advective velocity, $v_L$, with which the radiative energy density would be carried to account for the entire luminosity, $L$,
\begin{equation}\label{eq:pmn_vel}
v_{\rm L}(r) \equiv \frac{L}{4 \pi r^2 a T^4(r)},
\end{equation}
where $a$ is the radiation constant, and $r$ and $T$ are the local radius and temperature. 
To convert this advective velocity to a dimensionless number, we divide by the isothermal gas sound speed, $c_{\rm s} = \sqrt{P_{\rm gas}/\rho} = \sqrt{k_{\rm B} T/\mu m_{\rm p}}$, to create a pseudo-Mach number, $\pmn$,
\begin{equation}\label{eq:pmn}
    \pmn(r) \equiv \frac{v_{\rm L}(r)}{c_{\rm s}(r)} = \frac{L}{4 \pi a r^2 T^{4.5}(r)}\left(\frac{\mu m_{\rm p}}{k_{\rm B}}\right)^{1/2}, 
\end{equation}
where $\mu$ is the mean molecular weight, $k_B$ is the Boltzmann constant, and $m_p$ is the mass of a proton.
The choice of the isothermal gas sound speed arises because the largest fluctuations occur in regions where $\tau \ll \tau_{\rm c}$ implying photons are leaking out of fluid parcels rather than supporting them against compression \citep{Jiang2015}.
This leaking forces the gas pressure to be the only post-shock support, and thus the isothermal gas sound speed is a good choice for characterizing the amplitude of density fluctuations.
Despite the factor of $T^{-4.5}$, the fluctuations of $\pmn$ are still substantially smaller than those of density or radiative flux.

Figure~\ref{fig:pmn_v_r} shows $\pmn(r)$ for all of our 3D models, using $T(r) = \langle T(r) \rangle$.
Despite the wide variance in radial extent, all models cover similar ranges of $\pmn$. 
Additionally, each model shows two slope changes.
Deep in the envelope (around $\pmn \sim 5$), the turbulent convection begins to carry a flux comparable to radiative diffusion above the iron opacity peak causing a change in the temperature gradient and giving rise to a change in the gradient of $\pmn$.
Near the surface (around $\pmn \sim 10^4$), the optical depth drops below unity and the radiation freely streams, causing a second change in the gradient. 
Both gradient changes occur at different radial locations in all the models, however each occurs at a similar value of $\pmn$. 
As $\pmn$ is a more universal value than radius for diagnosing the level of fluctuation, it will be solely used for the remainder of this work.
The $\pmn$ values of the red horizontal lines in Figure~\ref{fig:pmn_v_r} are chosen to match the same colored lines in earlier figures.

\begin{figure}[t!]
\centering
\includegraphics[width=\linewidth]{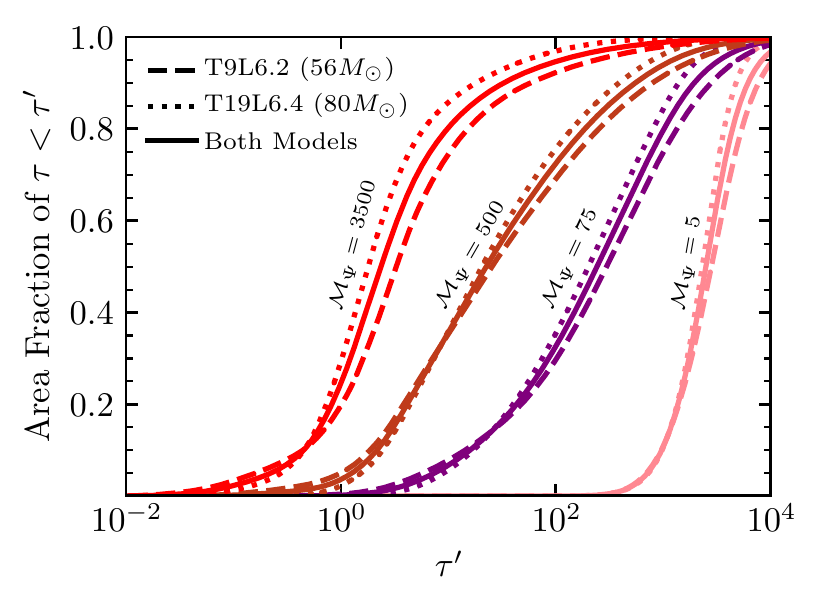}
\caption{Normalized cumulative sum of distributions of $\tau$ at four values of $\pmn$. 
Colors represent different choices of $\pmn$ and are labeled adjacently. 
Line styles correspond to $\tau$ distributions from different 3D models, T9L6.2 with $M_c = 56\,M_{\odot}$, and T19L6.4 with $M_c = 80\,M_{\odot}$.
Both models, consisting of the time average of 2800 snapshots, are combined and averaged to create a sample of 5600 snapshots shown by the solid line.  \label{fig:tau_cum_sum}}
\end{figure}

Figure~\ref{fig:tau_cum_sum} shows the cumulative distribution of the radial area fraction with $\tau$ less than a given value at four chosen values of $\pmn$, all above the iron opacity peak.
The three distributions plotted are generated from 2800 snapshots ($\approx 15$ days) of model T9L6.2 (dashed lines), 2800 snapshots ($\approx 15$ days) of model T19L6.4 (dotted lines), and the sum of the two, a 5600 snapshot sample (solid lines).
These snapshots were chosen to encompass the full range of luminosity variations observed in both models.
For both models considered, roughly 20$\%$ of the area at radii where $\pmn = 3500$ and 5$\%$ of the area where $\pmn = 500$ have optical depths to infinity of less than 1. 
These match the previously plotted red lines in Figures~\ref{fig:tau_v_r} and \ref{fig:other_v_r}.
Looking at the distributions themselves, the locations with $\pmn = 3500$ are not optically thick as the $\tau$ distribution rises sharply after $\tau=1$ and has 85$\%$ of area with $\tau < 10$.
The distributions corresponding to locations where $\pmn=500$ are less extreme with shallower slopes and only 50$\%$ of area having $\tau < 10$.
Despite the two models having different core masses, luminosities, radial extents, and effective temperatures, the optical depth distributions are very similar at locations with equal $\pmn$ values, confirming the selection of $\pmn$ as a good choice for an independent variable. 
It also provides an additional criterion for when the model starts to become optically thin.

\begin{figure*}[ht!]
\centering
\gridline{\fig{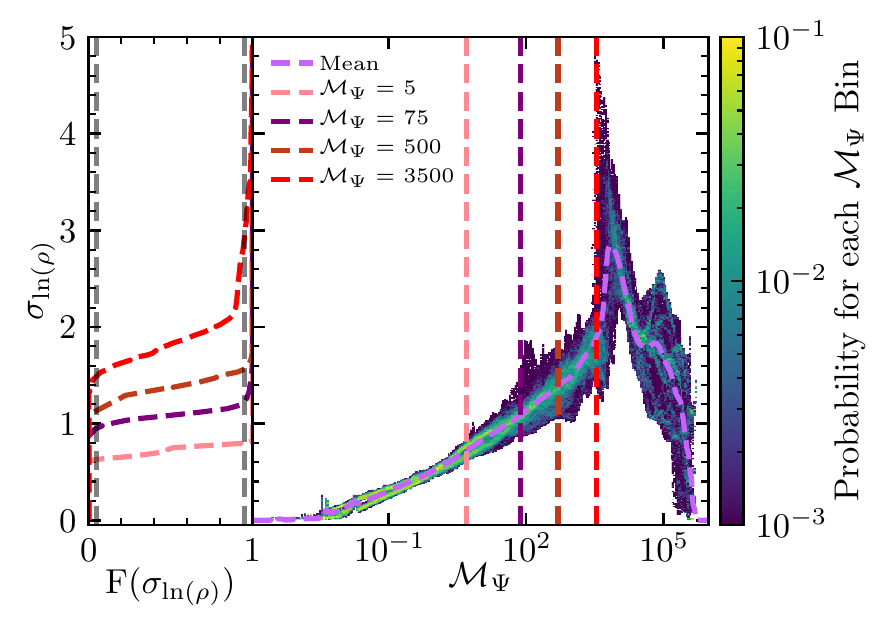}{0.45\linewidth}{(a)}
\fig{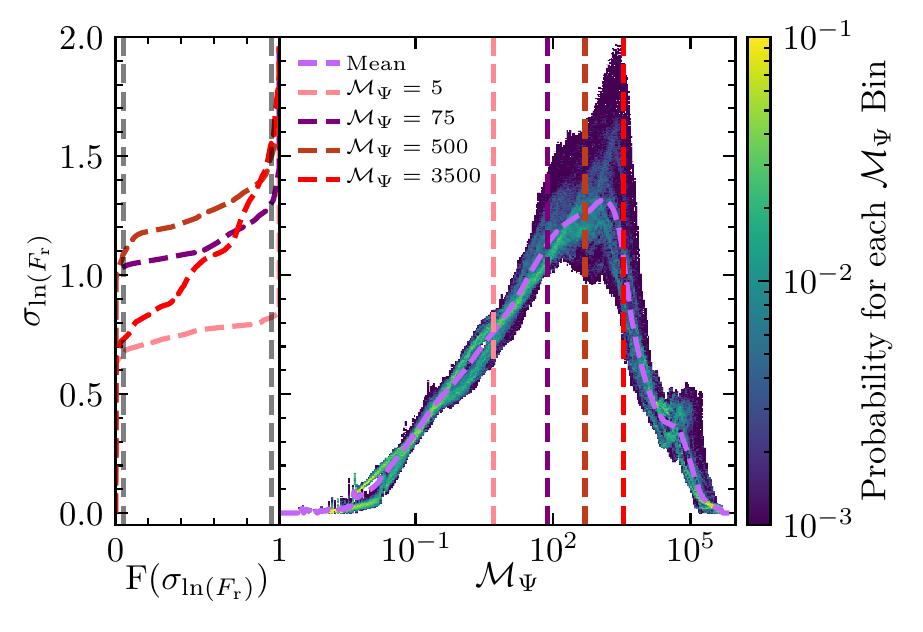}{0.45\linewidth}{(b)}
}
\gridline{\fig{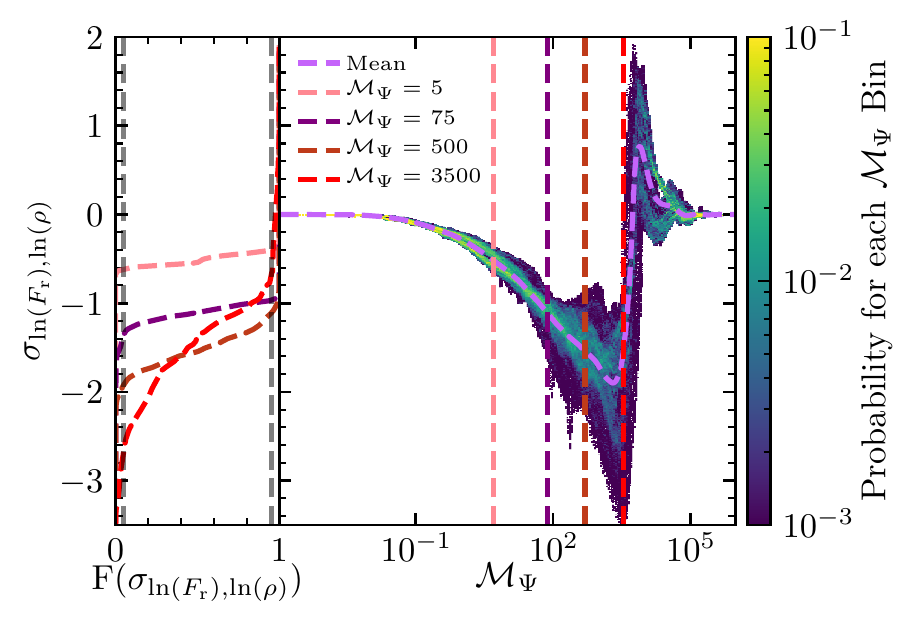}{0.45\linewidth}{(c)}
\begin{minipage}[b]{0.45\textwidth}
\caption{Standard deviation of (a) $\ln(\rho)$ and (b) $\ln(F_r)$, and (c) the covariance of $\ln(\rho)$ and $\ln(F_r)$ as a function of $\pmn$.
The right panel of each figure shows the distribution of the quantity as a function of $\pmn$ where the color represents the probability of each value for each $\pmn$ bin. 
These distributions are generated from 5600 3D snapshots, 2800 from model T9L6.2 and 2800 from model T19L6.4. 
The light purple-dashed line represents the mean of the distributions. 
The left panel shows the cumulative distribution function, $F(\sigma)$, of the quantity at chosen $\pmn$ values, $\pmn =$ 5, 75, 500, and 3500 (pink, purple, dark red, and bright red respectively).
The dashed, vertical grey lines show where the cumulative distribution is 5$\%$ and 95$\%$. \label{fig:sig_and_cov}}
\end{minipage}
}
\end{figure*}

\subsection{Variance and Covariance of Density and Radiative Flux} \label{sec:sig_cov}
The density and radiative flux follow log-normal distributions for the majority of the optically thick region of our models. 
As they are highly inversely correlated and affect the radiation pressure gradient (and hence the structure of the envelopes themselves) it is important to quantitatively characterize their standard deviations and covariance.
Figure~\ref{fig:sig_and_cov} shows the distributions of the standard deviations of $\ln(\rho)$, $\sigma_{\ln(\rho)}$, and $\ln(F_{\rm r})$, $\sigma_{\ln(F_{\rm r})}$, as well as their covariance, $\sigma_{\ln(F_{\rm r}),\ln(\rho)}$, as functions of $\pmn$.

The color represents the probability of observing a given value for the standard deviation or covariance at a certain $\pmn$, calculated from 5600 snapshots spanning $\approx 15$ days of two models (2800 from T9L6.2 and 2800 from T19L6.2).
For all quantities, the values are very constrained despite the large variety of stellar models.
Though two distributions can be distinguished at $\pmn \lesssim 10$, one from each model used, the distinction is neglected and we consider each variance to follow a single distribution over all values of $\pmn$.

The light purple-dashed lines represent the mean of the colored distributions.
The mean standard deviations of both $\ln(\rho)$ and $\ln(F_{\rm r})$, as well as the mean of their covariance, correspond to variations in $\rho$ and $F_{\rm r}$ larger than a factor of 2 at $\pmn = 10$.
\hlm{Moving outwards, the sound speed decreases, increasing the strength of the turbulent motion and thus increasing the variations} to a factor of 4 at $\pmn = 500$ where an optically thick assumption, like that used in equation~(\ref{eq:Pr_OG}), begins to degrade.
At $\pmn > 500$, the distributions of density and flux at fixed $\pmn$ start to deviate from log-normal and so despite the standard deviation of $\ln(\rho)$ and covariance approaching expected values (corresponding to factors of 10) these estimations should be taken less quantitatively.
In the optically thin region of the models, $\pmn > 3500$, the fluctuations cannot be described by a single log-normal distribution.
The characteristics of this region are strongly affected by the wind dynamics and are significantly time dependent which, combined with the fluctuations of the distributions from log-normal in shape, cause the lack of a clear expected value for the standard deviations and covariance. The physics of this region is beyond the scope of this work.

The left panel of each plot shows the cumulative distribution functions, $F(\sigma_{\ln(x)})$, of the standard deviation or covariance at the values of $\pmn$ specified by the colored vertical dashed lines in the right panels.
The cumulative distribution functions are defined as,
\begin{equation}
    F(\sigma_{\ln(x)}) = \int^{\sigma_{\ln(x)}}_{0} p(\sigma_{\ln(x)}') d\sigma_{\ln(x)}',
\end{equation}
where $p(\sigma_{\ln(x)}')$ is the probability density for that value of $\sigma_{\ln(x)}'$ at that value of pseudo-Mach number.
The vertical grey-dashed lines in the left panels show the 90$\%$ confidence intervals for the means. 
The cumulative distribution functions quantify the distributions that are shown qualitatively with color in the right panels, especially the low probability regions, shown in dark purple.
Though these excursions are large at high $\pmn$, such as $\pmn = 3500$, the majority of the probability lies in the central part of the distributions.
Typically the 90$\%$ confidence intervals span $\pm 0.5$, with higher values of $\pmn$'s having intervals of $\pm 1 $ relative to the mean value. 

In previous literature \citep[e.g.][]{Owocki2018}, density fluctuations occurring in turbulent media, typically clumpy stellar winds, are often characterized by the amplitude of over-densities, defined by $\frac{\langle \rho^2 \rangle}{\langle \rho \rangle^2}$. 
This definition of clumping is simply related to the standard deviation in the natural log of density, $\sigma_{\ln(\rho)}$.
For log-normal distributions, it can be shown that $\ln(\langle \rho \rangle) = \langle \ln(\rho) \rangle + \frac{\sigma^2_{\ln(\rho)}}{2}$ and $\ln(\langle \rho^2 \rangle) = 2\langle \ln(\rho) \rangle + 2\sigma_{\ln(\rho)}^2$.
Substituting these equation for the typical definition of over-densities in turbulent media yields,
\begin{equation}
    \sigma_{\ln(\rho)}^2 = \ln\left(\frac{\langle \rho^2 \rangle}{\langle \rho \rangle^2}\right).
\end{equation}
Hence squaring the distributions in Figure~\ref{fig:sig_and_cov}a reveals of the over-densities, or clumping factors, of the turbulent envelope.

\section{Calibrating the Impact of Fluctuations on Optically Thick Transport} \label{sec:psi}
The 3D models exhibit a strong correlation between density and radiation pressure which we must include in 1D models.
One dimensional models calculate the local radiation pressure gradient in optically thick regions, $(\nabla P_{\rm r})_{1D}$, using the diffusion equation,
\begin{equation} \label{eq:Pr_1D}
    (\nabla P_{\rm r})_{1D} = -\frac{1}{c} \langle F_{\rm r} \rangle \langle \kappa \rangle \langle \rho \rangle.
\end{equation}
This approach does not include the correlations of density, radiative flux, and opacity evident in 3D models, and thus does not agree with $\nabla P_{\rm r}$ from 3D models, which will satisfy
\begin{equation} \label{eq:Pr_3D}
    (\nabla P_{\rm r})_{3D} = -\frac{1}{c} \langle F_{\rm r}  \kappa  \rho \rangle.
\end{equation}
The opacity \hlm{is nearly constant} when compared to density or radiative flux and so we neglect the correlations with $\kappa$ and pull the average of $\kappa$ out of the main average of $\nabla P_{\rm r}$. 
\hlm{Extracting $\kappa$ from the average does not substantially affect the calculation of the radiation pressure; the relative error between $\langle \kappa \rangle \langle F_{\rm r} \rho \rangle$ and $\langle F_{\rm r} \kappa \rho \rangle$ is less than 0.1$\%$ within the range of $\pmn$ we are interested in.}
The ratio of the gradient from the 3D models, $\nabla P_{\rm r}$, and the 1D prediction of equation~(\ref{eq:Pr_1D}) then defines a new, dimensionless parameter, $\Psi$, that will account for the difference between the gradients,
\begin{equation} \label{eq:psi}
    \Psi \equiv \frac{\langle F_{\rm r} \rho \rangle}{\langle F_{\rm r} \rangle \langle \rho \rangle}.
\end{equation}
As $F_{\rm r}$ and $\rho$ are inversely correlated (detailed in Section~\ref{subsec:correlations}) $\Psi$ will never exceed unity.
The densest regions will have minimal radiative flux transport while the rarefied regions will have more.
The largest densities are multiplied by the smallest fluxes, and vice versa, in the numerator of equation~(\ref{eq:psi}), significantly reducing the average of the product when compared to the product of the independent averages.

By definition $\Psi$ is invariant to the means of $\rho$ or $F_{\rm r}$ allowing us to calculate it without using a specific model.
Using the standard deviations and covariance of $\rho$ and $F_{\rm r}$, shown in Figure~\ref{fig:sig_and_cov}, we synthesize a 2D probability distribution for density and radiative flux at each value of $\pmn$.
We generate $10^4$ mock values of both $\rho$ and $F_{\rm r}$ spanning five standard deviations evenly in log-space.
These values are collected to produce $10^8$ (density, radiative flux) pairs. Using the probability of each combination as weights, we calculate the averages of density, radiative flux, and their product and thus obtain values of $\Psi$.

\begin{figure}[t!]
\centering
\includegraphics[width=\linewidth]{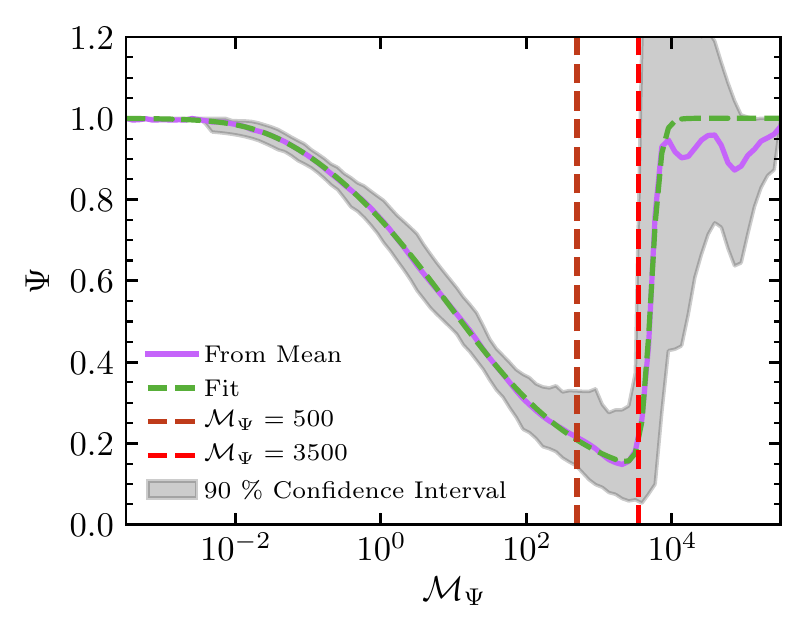}
\caption{Dependence of $\Psi$ on $\pmn$. 
The light purple line represents $\Psi$ as calculated from the mean of the variance distributions (light purple-dashed lines in Figure~\ref{fig:sig_and_cov}). 
The grey regions are the 90$\%$ confidence intervals from this mean.
The green-dashed line follows the functional fit for $\Psi$ given in equation~(\ref{eq:Psi_fit}). 
The vertical dashed lines are at $\pmn = 500$ (dark red) and $\pmn = 3500$ (bright red).
\label{fig:psi}}
\end{figure}

Figure~\ref{fig:psi} shows the calculated values of $\Psi$ as a function of $\pmn$.
The light purple line shows the mean values and the grey region corresponds to the 90$\%$ confidence intervals around the means.
The red vertical dashed lines show the region where an optically thick assumption degrades and where the variations of radiative flux and density deviate from log-normal distributions.
This deviation from log-normal distributions, due to the wind playing a dominant role in the dynamics, causes the uncertainty of $\Psi$ to substantially increase for $\pmn \gtrsim 3500$.
Proper characterization of this region is beyond the scope of this work as we choose to focus on the radiatively diffusive region where the radiative flux closely follows the diffusion equation (equation~(\ref{eq:Pr_3D})). 

The values of $\Psi$ become $\ll 1$ within the optically thick region, demonstrating that the 1D approximation of the radiation pressure gradient in equation~(\ref{eq:Pr_1D}) is a substantial over-estimate for these regions.
\hlm{The decrease in $\Psi$ arises from density and radiative flux fluctuations generated from the turbulence.
As the sound speed decreases with radius turbulent shocks become stronger causing larger density and radiative flux contrasts further our in the envelopes, at  higher $\pmn$ values.
Because the two quantities are highly anti-correlated, this reduces the average radiation pressure gradient of the 3D model, but does not substantially effect the expected values of density or radiative flux alone.
The reduction in the average of the radiation pressure gradient while leaving $\langle \rho \rangle$ and $\langle F_{\rm r} \rangle$ constant gives rise to the small values of $\Psi$.}

To simplify future calculations of $\Psi(\pmn)$, we fit the mean values using the product of a \hlm{hyperbolic tangent of} $\log_{10}(\pmn)$ and two Fermi functions.
The resulting formula is given by
\begin{equation} \label{eq:Psi_fit}
\begin{aligned}
   \Psi(\pmn) \approx 1 + &\Bigg\{ \Big(A\tanh{(B\log_{10}(\pmn) + C)} +D\Big) \times \\ 
    & f_{\rm Fermi}(-\log_{10}(\pmn), F_{1A},F_{1B}) \times \\
    & f_{\rm Fermi}(\log_{10}(\pmn),F_{2A},F_{2B}) \Bigg\} \\
   \text{where } f_{\rm Fermi}&(x, F_A, F_B) =\frac{1}{\exp\left(\frac{x + F_A}{F_B}\right) + 1},
\end{aligned}
\end{equation}
\hlm{and the fit coefficients are: $A=0.441$, $B=-0.533$, $C=0.525$, $D=-0.473$, $F_{1A}=-1.451$, $F_{1B}=0.465$, $F_{2A}=-3.715$, $F_{2B}=0.067$.}
This fitting function is shown by the green-dashed line in Figure~\ref{fig:psi}.
\hlm{Because of the large uncertainties and many of our assumptions breaking down, we have chosen to generate a fitting function that defaults to no modification to previous $(\nabla P_{\rm r})_{\rm 1D}$ calculation methods at large values of $\pmn$.
We understand that this choice may cause some difficulties when implemented in current 1D models and hence provide an alternative in Section~\ref{sec:1D_check}.}

\begin{figure}[t!]
\centering
\includegraphics[width=\linewidth]{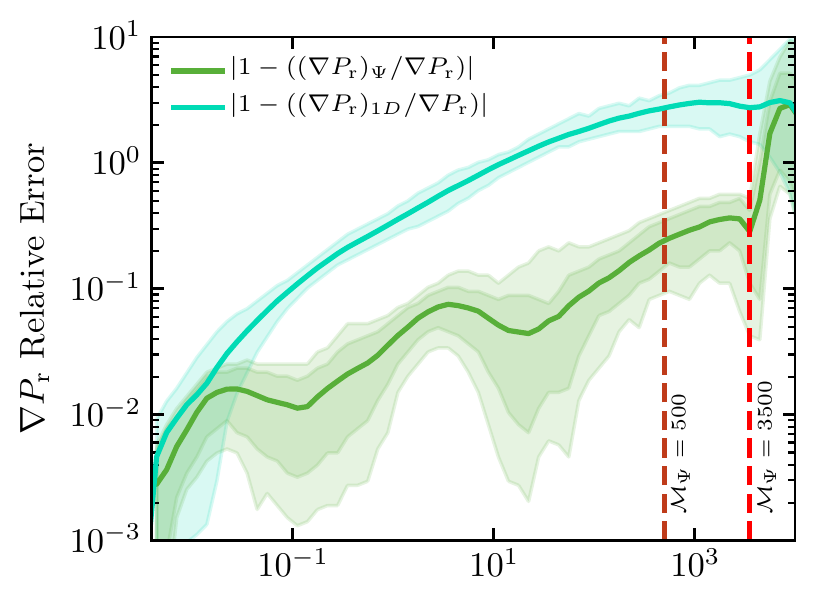}
\caption{Comparison of the fractional difference between the radiation pressure gradient calculated from the 3D model, $\nabla P_{\rm r}$, the 1D approximation, $(\nabla P_{\rm r})_{1D}$ (cyan), and the corrected $\left(\nabla P_{\rm r} \right)_{\Psi}$ (green).
The two distributions are generated using the same 5600 snapshots used to generate the distributions in Figure~\ref{fig:sig_and_cov}.
The solid lines show the means and the shaded regions represent the $70\%$ (darker) and $90\%$ (lighter) confidence intervals.
For the relative error of $(\nabla P_{\rm r})_{1D}$, only the $90\%$ confidence interval is plotted for clarity.
The vertical dashed lines are at $\pmn = 500$ (dark red) and $\pmn = 3500$ (bright red). \label{fig:train_comp}}
\end{figure}

To account for the correlation of radiative flux and density in the calculation of the 1D radiation pressure gradient in optically thick regions, we simply multiply $(\nabla P_{\rm r})_{1D}$ by $\Psi$ to get $(\nabla P_{\rm r})_{\Psi}$ or, 
\begin{equation} \label{eq:Pr_psi}
    (\nabla P_{\rm r})_{\Psi} \equiv -\frac{1}{c} \langle F_{\rm r} \rangle \langle \kappa \rangle \langle \rho \rangle \Psi.
\end{equation}
\hlm{This correction is only modifying the 1D radiation pressure gradient in optically thick regions. 
Until corrections accounting for turbulent pressure and the heat carried by convection are also incorporated, this definition of the radiation pressure gradient should not be implemented in a 1D stellar evolution code.
Characterizing turbulent pressure and the resulting convective flux will be addressed in future work.}

\begin{figure}[t!]
\centering
\includegraphics[width=\linewidth]{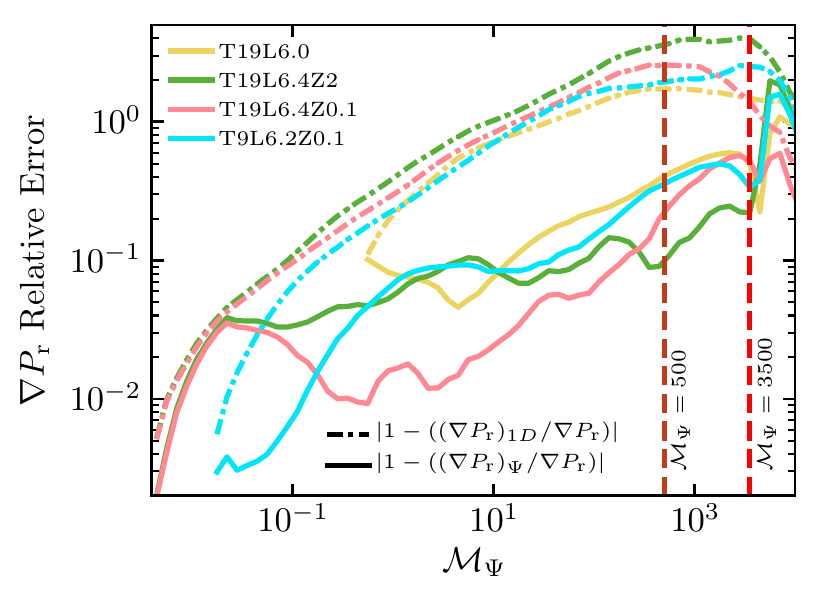}
\caption{
Comparison of the fractional difference between the radiation pressure gradient calculated from the 3D model, $\nabla P_{\rm r}$, the 1D approximation, $(\nabla P_{\rm r})_{1D}$ (dot-dashed), and the corrected 1D prescription, $\left(\nabla P_{\rm r} \right)_{\Psi}$ (solid) for models not used to generate $\Psi$.
Each line represents the mean of a distribution generated using more than 1000 snapshots ($> 5$ days of simulation time) from each model.
The models employ the same color coding as Figure~\ref{fig:pmn_v_r}.
The confidence intervals for each line are not plotted for clarity, but are similar to those shown in Figure~\ref{fig:train_comp}.
The vertical dashed lines are at $\pmn = 500$ (dark red) and $\pmn = 3500$ (bright red). \label{fig:mean_comp}}
\end{figure}

Figure~\ref{fig:train_comp} shows the relative error between the averaged 3D radiation pressure gradient, $\nabla P_{\rm r}$, the 1D approximation, $(\nabla P_{\rm r})_{1D}$ (cyan), and the corrected $(\nabla P_{\rm r})_{\Psi}$ (green), for all 5600 snapshots used to estimate the standard deviations and covariance of \hlm{ln(}$\rho$\hlm{)} and \hlm{ln(}$F_{\rm r}$\hlm{)} (see section~\ref{sec:sig_cov}).
The lines represent the mean and the shaded regions represent $70\%$ (darker) and $90\%$ (lighter) confidence intervals.
The fitting formula given in equation~(\ref{eq:Psi_fit}) determines values of $\Psi$ used for the calculation of $(\nabla P_{\rm r})_{\Psi}$.
The two distributions of relative error overlap for $\pmn \lesssim 10^{-2}$ where turbulent convection is not important and there is little variation in density and radiative flux.
However, at higher $\pmn$ values the relative errors differ substantially.
At $\pmn = 10$, $(\nabla P_{\rm r})_{1D}$ has a relative error of $100\%$, or a factor of 2 difference, while $(\nabla P_{\rm r})_{\Psi}$ only differs by $5\%$.
At $\pmn = 500$ (dark red-dashed line), where $5\%$ of the area has $\tau < 1$, $(\nabla P_{\rm r})_{1D}$ differs from $\nabla P_{\rm r}$ by nearly $300\%$, while $(\nabla P_{\rm r})_{\Psi}$ differs by only $22\%$.
The addition of $\Psi$ in the 1D approximation for the radiation pressure gradient results in an order of magnitude correction in relative error.
Above $\pmn = 3500$ (bright red-dashed line), $(\nabla P_{\rm r})_{\Psi}$ quickly approaches the same relative error as $(\nabla P_{\rm r})_{1D}$.
This is due to the fluctuations of radiative flux and density deviating from log-normal distributions. 
The transition of the probability distribution functions begins near $\pmn = 500$ (dark red vertical dashed line), and though the relative error of our new radiation pressure gradient prescription is promising when compared to the previous method, we caution the use of $(\nabla P_{\rm r})_{\Psi}$ above $\pmn = 500$.
Because of the sharp increase in the relative error of $(\nabla P_{\rm r})_{\Psi}$, we recommend $\pmn = 3500$ as the strict upper limit to the range of $\pmn$ over which $(\nabla P_{\rm r})_{\Psi}$ is an accurate approximation.

Implementing the same methods used to make Figure~\ref{fig:train_comp}, we now compare the relative error of 1D approximations to the radiation pressure gradient with and without our $\Psi$ parameter for the four envelope models listed in Table~\ref{tab:3D_models} that were not used to generate $\Psi$.
Figure~\ref{fig:mean_comp} shows the mean of each distribution generated using over 1000 snapshots ($> 5$ days) for each model.
These ranges of snapshots were chosen to encompass the majority of luminosity variations in the steady-state regions of these models.
These models are completely independent of those used in the fitting of our $\Psi$ parameter, making this a test of our modeling approach.
T19L6.0 (yellow lines) has a substantially less massive core and the others use the same core masses but with different metallicities.
As will be shown in future work, changes in metallicity affect the strength of the opacity peaks, and modify the turbulence resulting in different envelope structures. 
Because of these differences, comparing our prescription for the radiative pressure gradient in these models against the 3D results represents a test of our proposed method.

Deep in the models, the variance of density and radiative flux are small and so the difference is negligible, though the gradient estimate is improved with the addition of $\Psi$.
In all the models, the 1D radiation pressure gradient begins to diverge from the 3D results at the onset of the iron opacity peak, with the prescription including $\Psi$ staying closer to the 3D result.
In all the models, when the 1D approximation $(\nabla P_{\rm r})_{\rm 1D}$ reaches a relative error of $100\%$, the 1D estimate including $\Psi$ has a relative error of less than $17\%$, with all but T19L6.0 having an error of $\lesssim 10\%$.
Our prescription reaches a maximal relative error of $\approx 60\%$, though this occurs above $\pmn = 500$ where the optically thick assumption \hlm{and log-normal characterizations} begin to break down.
At $\pmn = 500$ the radiative pressure gradient prescription with $\Psi$ included has a relative error of $40\%$ or less, considerably lower than the $>100\%$ of the uncorrected formula.
Again $\pmn \approx 3500$ appears to be the limit above which our prescription for $\Psi$ starts to fail.
However, in this regions of higher $\pmn$ our prescription still outperforms the approximation without $\Psi$.

\section{Domain of Applicability in 1D Models and Future Work\label{sec:1D_check}}
Fundamentally, the turbulent convection that affects the radiation pressure gradient arises as the stellar luminosity approaches the Eddington luminosity, implying that the utilization of $\Psi$ to correct for radiative flux and density correlations is applicable to any astrophysical situation where any opacity peak arises in a near Eddington limited, radiation pressure dominated plasma. 
The envelopes of all stars with masses larger than $\approx 30\,M_{\odot}$ fulfill these requirements and thus 1D models of stars in this mass range will be affected by this correlation.
However, simply reducing the radiation pressure gradient in a 1D model is not a full correction, as the turbulent convection exerts substantial pressure and transports a fraction of the stellar luminosity.

For this reason, we cannot yet implement our $\Psi$ prescription in a modern 1D hydro-static stellar evolution code.
Instead, we show how substantial the impact of our new prescription would be when applied to models produced by the Modules for Experiments in Stellar Astrophysics \citep[MESA;][]{Paxton2011, Paxton2013, Paxton2015, Paxton2018, Paxton2019}.

\begin{figure}[ht!]
\centering
\includegraphics[width=\linewidth]{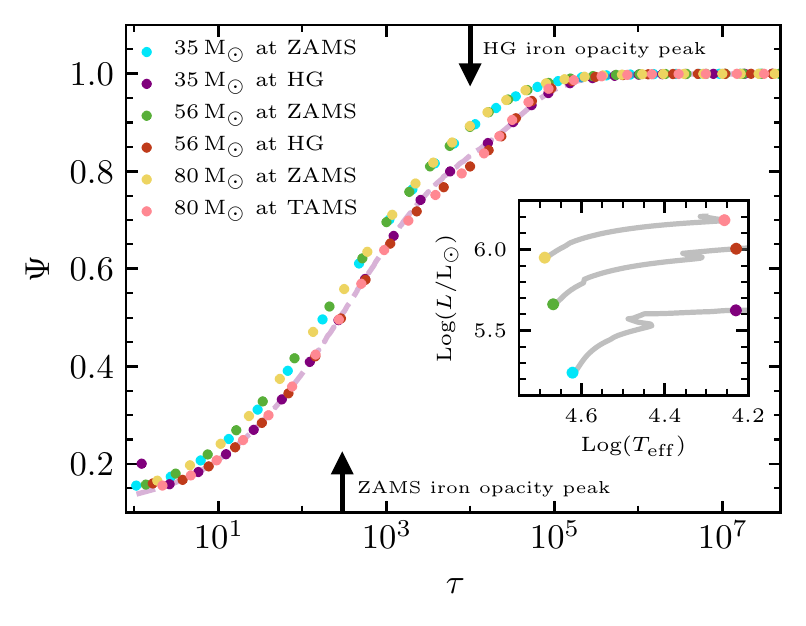}
\caption{Profiles of $\Psi$ versus optical depth for higher and lower $T_{\rm eff}$ models of $35\,$M$_{\odot}$, $56\,$M$_{\odot}$, and $80\,$M$_{\odot}$ solar metallicity stars.
The points \hlm{show the expected values of $\Psi$ for each model using equation~(\ref{eq:Psi_fit})} but do not represent the resolution of the models; rather they are spaced for clarity of viewing all profiles present.
The profiles are smooth between the plotted points.
\hlm{The light purple dashed line shows the full profile associated with the purple points using an alternative definition of $\Psi$ given by equation~(\ref{eq:new_Psi_fit}).
All $\Psi$ values are calculated in post-processing, after the models have been calculated.}
The black arrows indicated the location of the peak of the iron opacity peak for the hotter $T_{\rm eff}$, zero age main sequence (ZAMS) profiles (bottom) as well as the cooler, Hertzsprung gap (HG) and terminal age main sequence (TAMS) profiles (top).
Inset is a Hertsprung-Russell diagram of the MESA models' stellar evolution.
The grey lines show the stellar tracks. 
The colored points correspond to the luminosity, $L$, and effective temperature, $T_{\rm eff}$, of the models with $\Psi$ profiles plotted. \label{fig:1D_check}}
\end{figure}

Using mostly default values, we only set the initial mass and metallicity of the models, as well as setting {\sf okay$\_$to$\_$reduce$\_$grad$\_$T$\_$excess = .false.} to ensure we are only using classical mixing length theory.
With these settings, we modeled $35\,$M$_{\odot}$, $56\,$M$_{\odot}$, and $80\,$M$_{\odot}$ stars through their main sequence lifetimes.
These initial models simply reveal the domain of applicability of our new prescription.
Specifically, we looked at two types of models: hot models from the zero age main sequence (ZAMS), and cooler models from either the terminal age main sequence (TAMS) or the Hertzsprung gap (HG).
From these models, we calculated the value of $\Psi$ at each location.
These $\Psi$ profiles are plotted versus $\tau$ in Figure~\ref{fig:1D_check}, and show that significant reductions in the radiation pressure gradient (small values of $\Psi$) are expected near the surface of all the models considered. 
Looking at the $\Psi$ profiles, there are clearly two distributions; one for the hotter $T_{\rm eff}$ stars and another for the lower $T_{\rm eff}$ stars (see inset HR diagram of Figure~\ref{fig:1D_check}).
This is due to a difference in both the location, shown by the two black arrows, and shape of their iron opacity peaks.
The lower $T_{\rm eff}$ models have a deeper iron opacity peak along with a steeper rise of the peak, starting at $\tau \approx 10^5$. 
This steep rise causes the initial sharper decrease in $\Psi$ at $\tau \approx 10^4$. 
The higher $T_{\rm eff}$ models have the iron opacity peak substantially closer to the surface and a much shallower rise, starting as deep as $\tau \approx 5\times 10^4$ and peaking at $\tau \approx 300$.
In all the profiles, our prescription would recommend reducing the radiation pressure gradient by over $80\%$ at the surface. 
The inclusion of this prescription will have profound effect on the temperature gradient, changing the nature of convection in these 1D envelopes and substantially affecting the evolution of these models in the HR diagram.

\hlm{Looking at the purple points in Figure~\ref{fig:1D_check}, we notice an upturn near $\tau \approx 1$ as a result from our choice of convention when generating equation~(\ref{eq:Psi_fit}).
As this upturn occurs quickly and near the surface of the models, it may cause some undesirable effects when implemented in 1D models.
It could potentially require stricter surface boundary conditions as well as higher resolutions to resolve the large slope change accurately.
Because the upturn in $\Psi$ is likely a product of poorly constrained approximations in this region, we suggest an alternative fitting equation that will alleviate these problem.
The generate the new formula, we remove the second Fermi function from equation~(\ref{eq:Psi_fit}):}
\begin{equation} \label{eq:new_Psi_fit}
\begin{aligned}
    \Psi(\pmn) \approx 1 + &\Bigg\{ \Big(A\tanh{(B\log_{10}(\pmn) + C)} +D\Big) \times \\ 
    & f_{\rm Fermi}(-\log_{10}(\pmn), F_{1A},F_{1B}) \Bigg\}
\end{aligned}
\end{equation}
\hlm{where the constants are the same as those in equation~(\ref{eq:Psi_fit}).
The resulting $\Psi$ profile is shown by the light purple line in Figure~\ref{fig:1D_check} and can be directly compared to the purple points.
The removal of the second Fermi function does not affect the majority of the profile, however near the surface of the model the slope does not change sign.
Instead, the new fitting function of $\Psi$ approaches $\approx 0.1$ at $\pmn \approx 10^6$. 
We are not claiming either equation~(\ref{eq:Psi_fit}) or equation~(\ref{eq:new_Psi_fit}) will work well in this region ($\pmn \gtrsim 3500$); we are only suggesting equation~(\ref{eq:new_Psi_fit}) might allow 1D models to converge more easily.}

\section{Conclusion \label{sec:conclusion}}
We quantified the large density and radiative flux fluctuations and their induced modification to the radiative pressure gradient needed to carry a certain radiative luminosity in optically thick massive star envelopes.
As seen in the recent 3D models of \citet{Jiang2018}, turbulent velocities drive shocks and very large density and radiative flux fluctuations (see Figure~\ref{fig:other_v_r}) that substantially modify the nature of radiative transport even at $\tau \gg 10$.
We showed that the density and radiative flux follow log-normal distributions throughout the optically thick region of our models allowing us to accurately quantify their variations by a pseudo-Mach number, $\pmn$.
We showed that we can successfully quantify the variance and covariance of $\rho$ and $F_{\rm r}$ with this single number at every location in thousands of snapshots from two distinct 3D models.
The choice of $\pmn$ arose from comparing the convective velocity needed to carry the heat to the isothermal gas sound speed. 
The isothermal gas sound speed was used because, in the regions of large density fluctuations, the optical depth is low enough the photons cannot prevent the gas from compressing, leaving the gas pressure as the only support.
After quantifying the fluctuations and their correlations as a function of $\pmn$, we derived an effective 1D implementation that yields $\lesssim 30\%$ agreement with the 3D $\nabla P_{\rm r}$ \hlm{in optically thick regions of the models}.
This prescription is described by a local quantity, $\pmn(L, T, r)$, and a fitting formula, equation~(\ref{eq:Psi_fit}), that will allow simple incorporation into future 1D models.\footnote{\hlm{Our fitting formula should only be trusted for $\pmn < 3500$.}}

Fundamentally, the turbulent convection that affects the radiation pressure gradient arises as a luminosity approaches the Eddington luminosity, implying that our correction will be applicable to other astrophysical situations where an opacity peak arises in a near Eddington limited, radiation pressure dominated plasma.
Recent work \citep{Jiang2016, Jiang2020} has shown the iron opacity peak modifies the structure and instigates convection throughout accretion disks around active galactic nuclei (AGN) suggesting our prescription could be useful in future 1D modeling of AGN evolution.
Additionally, the envelopes of all stars with masses larger than $\sim 30\,M_{\odot}$ fulfill these requirements and thus 1D models of stars in this mass range will be affected by this correlation.  
However, simply reducing the radiation pressure gradient is not enough to implement a new 1D model.
Rather, we must also account for the impact of two other aspects of turbulent convection, the substantial pressure and the ability to transport an appreciable fraction of the stellar luminosity.
This will be part of the focus of future efforts to improve 1D modelling of massive stars using physically motivated prescriptions based on locally defined parameters.

\acknowledgments

We thank Omer Blaes, Bill Paxton, and Benny Tsang for many helpful conversations and comments.
\hlm{We also thank the anonymous referee for their helpful and timely feedback.}
This research was supported in part by the NASA ATP grant ATP-80NSSC18K0560, by the National Science Foundation through grant PHY 17-48958 at the KITP and benefited from interactions that were funded by the Gordon and Betty Moore Foundation through Grant GBMF5076.
An award of computer time was provided by the Innovative and Novel Computational Impact on Theory and Experiment (INCITE) programme.
This research used resources of the Argonne Leadership Computing Facility and National Energy Research Scientific Computing Center, which are DOE Offices of Science User Facility supported under contract DE-AC02-06CH11357 and DE-AC02-05CH11231.
Resources supporting this work were also provided by the NASA High-End Computing (HEC) programme through the NASA Advanced Supercomputing (NAS) Division at Ames Research Center. 
We acknowledge support from the Center for Scientific Computing from the CNSI, MRL: an NSF MRSEC (DMR-1720256) and NSF CNS-1725797.
The Flatiron Institute is supported by the Simons Foundation.

\software{Matplotlib \citep{Hunter2007},
        NumPy \citep{numpy},
        SciPy \citep{scipy},
          MESA \citep{Paxton2011, Paxton2013, Paxton2015, Paxton2018, Paxton2019}
          }

\newpage
\bibliography{bib}{}

\begin{thebibliography}{}
\expandafter\ifx\csname natexlab\endcsname\relax\def\natexlab#1{#1}\fi
\providecommand{\url}[1]{\href{#1}{#1}}
\providecommand{\dodoi}[1]{doi:~\href{http://doi.org/#1}{\nolinkurl{#1}}}
\providecommand{\doeprint}[1]{\href{http://ascl.net/#1}{\nolinkurl{http://ascl.net/#1}}}
\providecommand{\doarXiv}[1]{\href{https://arxiv.org/abs/#1}{\nolinkurl{https://arxiv.org/abs/#1}}}

\bibitem[{{Bromm} \& {Larson}(2004)}]{Bromm2004}
{Bromm}, V., \& {Larson}, R.~B. 2004, \araa, 42, 79,
  \dodoi{10.1146/annurev.astro.42.053102.134034}

\bibitem[{{Cox} \& {Giuli}(1968)}]{Cox1968}
{Cox}, J.~P., \& {Giuli}, R.~T. 1968, {Principles of stellar structure}

\bibitem[{{Crowther}(2007)}]{Crowther2007}
{Crowther}, P.~A. 2007, \araa, 45, 177,
  \dodoi{10.1146/annurev.astro.45.051806.110615}

\bibitem[{{Davis} {et~al.}(2014){Davis}, {Jiang}, {Stone}, \&
  {Murray}}]{Davis2014}
{Davis}, S.~W., {Jiang}, Y.-F., {Stone}, J.~M., \& {Murray}, N. 2014, \apj,
  796, 107, \dodoi{10.1088/0004-637X/796/2/107}

\bibitem[{Davis {et~al.}(2012)Davis, Stone, \& Jiang}]{Davis2012}
Davis, S.~W., Stone, J.~M., \& Jiang, Y.-F. 2012, The Astrophysical Journal
  Supplement Series, 199, 9, \dodoi{10.1088/0067-0049/199/1/9}

\bibitem[{{Farmer} {et~al.}(2016){Farmer}, {Fields}, {Petermann}, {Dessart},
  {Cantiello}, {Paxton}, \& {Timmes}}]{Farmer2016}
{Farmer}, R., {Fields}, C.~E., {Petermann}, I., {et~al.} 2016, \apjs, 227, 22,
  \dodoi{10.3847/1538-4365/227/2/22}

\bibitem[{{Gr{\"a}fener} {et~al.}(2012){Gr{\"a}fener}, {Owocki}, \&
  {Vink}}]{Grafener2012}
{Gr{\"a}fener}, G., {Owocki}, S.~P., \& {Vink}, J.~S. 2012, \aap, 538, A40,
  \dodoi{10.1051/0004-6361/201117497}

\bibitem[{{Heger} {et~al.}(2003){Heger}, {Fryer}, {Woosley}, {Langer}, \&
  {Hartmann}}]{Heger2003}
{Heger}, A., {Fryer}, C.~L., {Woosley}, S.~E., {Langer}, N., \& {Hartmann},
  D.~H. 2003, \apj, 591, 288, \dodoi{10.1086/375341}

\bibitem[{Hunter(2007)}]{Hunter2007}
Hunter, J.~D. 2007, Computing in Science \& Engineering, 9, 90,
  \dodoi{10.1109/MCSE.2007.55}

\bibitem[{{Jiang} \& {Blaes}(2020)}]{Jiang2020}
{Jiang}, Y.-F., \& {Blaes}, O. 2020, arXiv e-prints, arXiv:2006.08657.
\newblock \doarXiv{2006.08657}

\bibitem[{Jiang {et~al.}(2015)Jiang, Cantiello, Bildsten, Quataert, \&
  Blaes}]{Jiang2015}
Jiang, Y.-F., Cantiello, M., Bildsten, L., Quataert, E., \& Blaes, O. 2015, The
  Astrophysical Journal, 813, 74, \dodoi{10.1088/0004-637x/813/1/74}

\bibitem[{Jiang {et~al.}(2018)Jiang, Cantiello, Bildsten, Quataert, Blaes, \&
  Stone}]{Jiang2018}
Jiang, Y.-F., Cantiello, M., Bildsten, L., {et~al.} 2018, Nature, 561, 498,
  \dodoi{10.1038/s41586-018-0525-0}

\bibitem[{Jiang {et~al.}(2016)Jiang, Davis, \& Stone}]{Jiang2016}
Jiang, Y.-F., Davis, S.~W., \& Stone, J.~M. 2016, The Astrophysical Journal,
  827, 10, \dodoi{10.3847/0004-637x/827/1/10}

\bibitem[{Jiang {et~al.}(2014)Jiang, Stone, \& Davis}]{Jiang2014}
Jiang, Y.-F., Stone, J.~M., \& Davis, S.~W. 2014, The Astrophysical Journal
  Supplement Series, 213, 7, \dodoi{10.1088/0067-0049/213/1/7}

\bibitem[{{Joss} {et~al.}(1973){Joss}, {Salpeter}, \& {Ostriker}}]{Joss1973}
{Joss}, P.~C., {Salpeter}, E.~E., \& {Ostriker}, J.~P. 1973, \apj, 181, 429,
  \dodoi{10.1086/152060}

\bibitem[{{Kennicutt}(2005)}]{Kennicutt2005}
{Kennicutt}, R.~C. 2005, in IAU Symposium, Vol. 227, Massive Star Birth: A
  Crossroads of Astrophysics, ed. R.~{Cesaroni}, M.~{Felli}, E.~{Churchwell},
  \& M.~{Walmsley}, 3--11, \dodoi{10.1017/S1743921305004308}

\bibitem[{{K{\"o}hler} {et~al.}(2015){K{\"o}hler}, {Langer}, {de Koter}, {de
  Mink}, {Crowther}, {Evans}, {Gr{\"a}fener}, {Sana}, {Sanyal}, {Schneider}, \&
  {Vink}}]{Kohler2015}
{K{\"o}hler}, K., {Langer}, N., {de Koter}, A., {et~al.} 2015, \aap, 573, A71,
  \dodoi{10.1051/0004-6361/201424356}

\bibitem[{{Krumholz} \& {Thompson}(2012)}]{Krumholz2012}
{Krumholz}, M.~R., \& {Thompson}, T.~A. 2012, \apj, 760, 155,
  \dodoi{10.1088/0004-637X/760/2/155}

\bibitem[{{Krumholz} \& {Thompson}(2013)}]{Krumholz2013}
---. 2013, \mnras, 434, 2329, \dodoi{10.1093/mnras/stt1174}

\bibitem[{{Maeder}(1987)}]{Maeder1987}
{Maeder}, A. 1987, \aap, 173, 247

\bibitem[{{Maeder} {et~al.}(2012){Maeder}, {Georgy}, {Meynet}, \&
  {Ekstr{\"o}m}}]{Maeder2012}
{Maeder}, A., {Georgy}, C., {Meynet}, G., \& {Ekstr{\"o}m}, S. 2012, \aap, 539,
  A110, \dodoi{10.1051/0004-6361/201118328}

\bibitem[{Oliphant(2006--)}]{numpy}
Oliphant, T. 2006--, {NumPy}: A guide to {NumPy}, USA: Trelgol Publishing.
\newblock \url{http://www.numpy.org/}

\bibitem[{{Owocki}(2015)}]{Owocki2015}
{Owocki}, S.~P. 2015, Astrophysics and Space Science Library, Vol. 412,
  {Instabilities in the Envelopes and Winds of Very Massive Stars}, ed. J.~S.
  {Vink}, 113, \dodoi{10.1007/978-3-319-09596-7_5}

\bibitem[{{Owocki} \& {Sundqvist}(2018)}]{Owocki2018}
{Owocki}, S.~P., \& {Sundqvist}, J.~O. 2018, \mnras, 475, 814,
  \dodoi{10.1093/mnras/stx3225}

\bibitem[{{Paxton} {et~al.}(2011){Paxton}, {Bildsten}, {Dotter}, {Herwig},
  {Lesaffre}, \& {Timmes}}]{Paxton2011}
{Paxton}, B., {Bildsten}, L., {Dotter}, A., {et~al.} 2011, \apjs, 192, 3,
  \dodoi{10.1088/0067-0049/192/1/3}

\bibitem[{{Paxton} {et~al.}(2013){Paxton}, {Cantiello}, {Arras}, {Bildsten},
  {Brown}, {Dotter}, {Mankovich}, {Montgomery}, {Stello}, {Timmes}, \&
  {Townsend}}]{Paxton2013}
{Paxton}, B., {Cantiello}, M., {Arras}, P., {et~al.} 2013, \apjs, 208, 4,
  \dodoi{10.1088/0067-0049/208/1/4}

\bibitem[{{Paxton} {et~al.}(2015){Paxton}, {Marchant}, {Schwab}, {Bauer},
  {Bildsten}, {Cantiello}, {Dessart}, {Farmer}, {Hu}, {Langer}, {Townsend},
  {Townsley}, \& {Timmes}}]{Paxton2015}
{Paxton}, B., {Marchant}, P., {Schwab}, J., {et~al.} 2015, \apjs, 220, 15,
  \dodoi{10.1088/0067-0049/220/1/15}

\bibitem[{{Paxton} {et~al.}(2018){Paxton}, {Schwab}, {Bauer}, {Bildsten},
  {Blinnikov}, {Duffell}, {Farmer}, {Goldberg}, {Marchant}, {Sorokina},
  {Thoul}, {Townsend}, \& {Timmes}}]{Paxton2018}
{Paxton}, B., {Schwab}, J., {Bauer}, E.~B., {et~al.} 2018, \apjs, 234, 34,
  \dodoi{10.3847/1538-4365/aaa5a8}

\bibitem[{{Paxton} {et~al.}(2019){Paxton}, {Smolec}, {Schwab}, {Gautschy},
  {Bildsten}, {Cantiello}, {Dotter}, {Farmer}, {Goldberg}, {Jermyn}, {Kanbur},
  {Marchant}, {Thoul}, {Townsend}, {Wolf}, {Zhang}, \& {Timmes}}]{Paxton2019}
{Paxton}, B., {Smolec}, R., {Schwab}, J., {et~al.} 2019, \apjs, 243, 10,
  \dodoi{10.3847/1538-4365/ab2241}

\bibitem[{{Rosdahl} \& {Teyssier}(2015)}]{Rosdahl2015}
{Rosdahl}, J., \& {Teyssier}, R. 2015, \mnras, 449, 4380,
  \dodoi{10.1093/mnras/stv567}

\bibitem[{{Sanyal} {et~al.}(2015){Sanyal}, {Grassitelli}, {Langer}, \&
  {Bestenlehner}}]{Sanyal2015}
{Sanyal}, D., {Grassitelli}, L., {Langer}, N., \& {Bestenlehner}, J.~M. 2015,
  \aap, 580, A20, \dodoi{10.1051/0004-6361/201525945}

\bibitem[{{Smith}(2014)}]{Smith2014}
{Smith}, N. 2014, \araa, 52, 487, \dodoi{10.1146/annurev-astro-081913-040025}

\bibitem[{Stone {et~al.}(2020)Stone, Tomida, White, \& Felker}]{Stone2020}
Stone, J.~M., Tomida, K., White, C.~J., \& Felker, K.~G. 2020, The Athena++
  Adaptive Mesh Refinement Framework: Design and Magnetohydrodynamic Solvers.
\newblock \doarXiv{2005.06651}

\bibitem[{{Stothers} \& {Chin}(1979)}]{Stothers1979}
{Stothers}, R., \& {Chin}, C.~W. 1979, \apj, 233, 267, \dodoi{10.1086/157388}

\bibitem[{{Tsang} \& {Milosavljevi{\'c}}(2015)}]{Tsang2015}
{Tsang}, B. T.~H., \& {Milosavljevi{\'c}}, M. 2015, \mnras, 453, 1108,
  \dodoi{10.1093/mnras/stv1707}

\bibitem[{{Tsang} \& {Milosavljevi{\'c}}(2018)}]{Tsang2018}
---. 2018, \mnras, 478, 4142, \dodoi{10.1093/mnras/sty1217}

\bibitem[{{Virtanen} {et~al.}(2020){Virtanen}, {Gommers}, {Oliphant},
  {Haberland}, {Reddy}, {Cournapeau}, {Burovski}, {Peterson}, {Weckesser},
  {Bright}, {van der Walt}, {Brett}, {Wilson}, {Jarrod Millman}, {Mayorov},
  {Nelson}, {Jones}, {Kern}, {Larson}, {Carey}, {Polat}, {Feng}, {Moore}, {Vand
  erPlas}, {Laxalde}, {Perktold}, {Cimrman}, {Henriksen}, {Quintero}, {Harris},
  {Archibald}, {Ribeiro}, {Pedregosa}, {van Mulbregt}, \&
  {Contributors}}]{scipy}
{Virtanen}, P., {Gommers}, R., {Oliphant}, T.~E., {et~al.} 2020, Nature
  Methods, 17, 261, \dodoi{https://doi.org/10.1038/s41592-019-0686-2}

\end{thebibliography}
\bibliographystyle{aasjournal}



\end{CJK*}
\end{document}